\documentclass{vgtc}                          

\ifpdf
  \pdfoutput=1\relax                   
  \usepackage{graphicx}                
  \DeclareGraphicsExtensions{.pdf,.png,.jpg,.jpeg} 
\else
  \usepackage{graphicx}                
  \DeclareGraphicsExtensions{.eps}     
\fi%

\graphicspath{{figures/}{pictures/}{images/}{./}} 

\usepackage{microtype}                 
\PassOptionsToPackage{warn}{textcomp}  
\usepackage{textcomp}                  
\usepackage{mathptmx}                  
\usepackage{times}                     
\usepackage{cite}                      
\usepackage{tabu}                      
\usepackage{booktabs}                  

\newcommand{\figref}[1]{Fig.~\ref{fig:#1}}

\onlineid{0}

\vgtccategory{Tool}

\vgtcinsertpkg

\title{BrainViewer: interacting with spatial connectome data at the mesoscale}

\author{Seth Daetwiler\thanks{e-mail: daetwis@wwu.edu} \\
\parbox{1.495in}{\scriptsize \centering
Computer Science\\
Western Washington University
     }
\and Angus Read\thanks{e-mail: reada2@wwu.edu}\\ %
\parbox{1.495in}{\scriptsize \centering
Computer Science\\
Western Washington University
     }
     \and Jessica Stillwell\thanks{e-mail: stillwj3@wwu.edu}\\ %
\parbox{1.495in}{\scriptsize \centering
Computer Science\\
Western Washington University
     }
     \and Kameron Decker Harris\thanks{e-mail: harri267@wwu.edu, corresponding author}\\ %
\parbox{1.495in}{\scriptsize \centering
Computer Science\\
Western Washington University
     }     
     }

\abstract{
Scientists construct connectomes, comprehensive descriptions of neuronal connections across a brain, in order to better understand and model brain function.
Interactive visualizations of these pathways would enable exploratory analysis of such information flows. 
Current tools can be used to see individual tracing experiments which are used to build mesoscale connectomes of the mouse brain, but not the brain network itself.
We present a connectivity visualization program called BrainViewer, which we use with a high-resolution mouse cortical connectome.
This has the ability to display connectomes from other datasets when they become available and compare spatial connectivity across multiple brain structures.
Our tool, optimized for speed and portability, presents a GUI visualization in 2-D top view and flatmap projections, allowing users to select and explore the connections of every source voxel to everywhere else in the cortex.
Anatomists and other neuroscientists will find BrainViewer useful for building understanding beyond the known topography of cortical connectivity.
} 

\CCScatlist{
  \CCScatTwelve{Computing methodologies}{Modeling and simulation}{}{};
  \CCScatTwelve{Human-centered computing}{Interactive systems and tools}{}{};
  \CCScatTwelve{Human-centered computing}{Visualization systems and tools}{}{};
  \CCScatTwelve{Human-centered computing}{Visualization application domains}{}{};
  \CCScatTwelve{Software and its engineering}{Software creation and management}{}{};
}

\begin{document}


\firstsection{Introduction}

\maketitle

Connectivity patterns in the brain reveal important information about brain function.
Differences in brain-wide connectivity patterns, or {\em connectomes}, 
are at the root of some functional and behavioral differences across animals.
Improved connectomic knowledge could lead to new treatments for disease, improved brain-machine interfaces, and influence the development of brain-inspired computing and artificial intelligence.

Topography refers to the functional spatial maps across the brain.
For instance, it is known that many sensory systems maintain maps of sensory 
space which are transformed between brain regions. 
Retinotopy--the mapping of visual space onto visual regions of cortex--is 
one example of this and was the original inspiration for deep convolutional 
neural networks which have revolutionized computer vision \cite{fukushima1980,lecun2015}.
It is now evident that retinotopy arises from topographic patterning of connections between brain areas \cite{zhuang2017}, however such data have been difficult to visualize.

Modern neuronal tracing methods allow researchers to trace the major pathways in the brain.
Such tracing experiments have been conducted on the mouse \cite{oh_mesoscale_2014, gamanut_mouse_2018, kuan_neuroinformatics_2015, zingg_neural_2014}, fly \cite{shih_connectomics-based_2015}, marmosets \cite{woodward_brainminds_2018, Hira_Modules, majka_towards_2016, watakabe_connectional_2021}, rat \cite{bota_gene_2003} and macaque \cite{markov_weighted_2014}.
Results of these experiments have then been used to build brain models at the mesoscale, i.e.\ showing the connections between areas containing on the order of thousands of neurons. 
Oh et al.\ used the anterograde tracer adeno-associated virus
expressing green fluorescent protein to track connections in the mouse brain \cite{oh_mesoscale_2014}. 
After injecting mice with the virus, each brain is then sliced, imaged and processed to reveal a volumetric image of fluorescence.
In the dataset we use from the Allen Institute for Brain Science, areas are defined using a common coordinate framework and individual experiments are
mapped into this common voxel space \cite{wang_allen_2020}.
Many of these experiments can be combined into a weighted connectivity matrix which stores projection strength between source and target regions \cite{oh_mesoscale_2014,knox_high-resolution_2018}, 
and individual tracing experiments can be visualized in 3D using the BrainExplorer tool \cite{lau_exploration_2008}.
Most connectome work has focused on measuring brain connectivity at the 
level of regions and various tools exist 
to display such regional connectivity.
New methods have been developed to estimate spatially-resolved connectivity at the resolution of voxels---discrete locations within the brain volume \cite{harris2016,knox_high-resolution_2018}.

The closest work to our own is the web-based
BrainModules from R.\ Hira \cite{Hira_Modules}
that displays functional and structural connectivity patterns from marmoset and mouse in the browser.
However, these are only available at low display resolution and the code is undocumented and unmaintained.

\begin{figure}[t!]
 \centering
 \includegraphics[width=\linewidth]{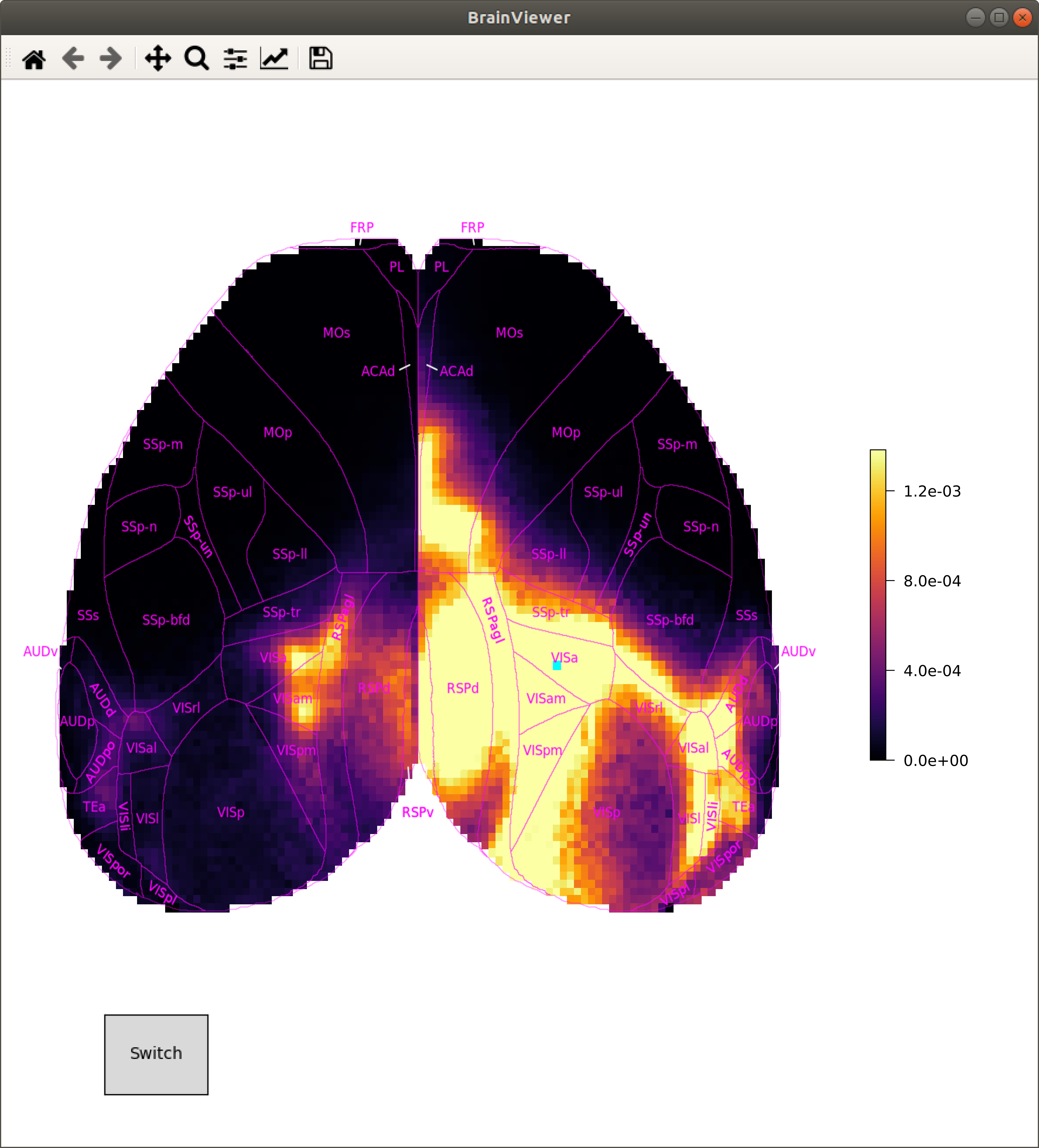}
  \caption{BrainViewer's GUI allows users to interact with connectome data.
  Here we show the mouse voxel connectome from \cite{knox_high-resolution_2018}.
  The source voxel (cyan) is seen in the VISa region and projection strength from that voxel is shown by the hot colors. 
  Cortical region boundaries and labels are overlaid in magenta. 
  The user may switch to a flatmap view with the button, navigate by keyboard, 
  and save resulting figures.}
  \label{fig:gui}
\end{figure}

We present a tool {\em BrainViewer} to display and interact with high resolution spatial brain connectivity.
We have built it to work with the mouse voxel connectome from 
\cite{knox_high-resolution_2018},
but it is portable and can be used across datasets
with a number of capabilities described in Section~\ref{sec:tool}.
The user may click on any location in right hemisphere of the mouse cortex and see the outgoing projections of neurons from that location.
Our tool provides a 2-D visualization of the connectivity in two projections, flatmap and top view; the relative advantages of each are discussed in the methods, Section~\ref{sec:methods}.
Switching between the two projections allows for easy comparison,
and the user may also navigate the source voxel using their keyboard.
This interactive visualization of a spatial connectome enhances the understanding of the brain’s information processing patterns and allow researchers to generate new and interesting topographic anatomical hypotheses.

\section{BrainViewer: a connectome visualizer tool}
\label{sec:tool}

Our tool allows users to navigate and explore the estimated mouse cortical connectivity
from Knox et al.\ \cite{knox_high-resolution_2018}
in a GUI environment
(see \figref{gui}).
The code is available at \url{https://github.com/glomerulus-lab/brainviewer}.
We include the following features:
\begin{itemize}
    \item A colormap annotating the strength of connectivity from the source voxel to any given voxel in the plot. 
    \item Click-to-plot allows users to select any voxel on the right hemisphere of either the flatmap or top view, visualizing the connectivity from that source voxel.
    \item Arrow key navigation which enables users to carefully move the source voxel across the cortex. 
    \item Switching between the top view and flatmap while maintaining the source voxel, allowing for visualization of connectivity in regions of the cortex obscured
    when viewed from above.
\end{itemize}

\subsection{Exploratory interaction capabilities}
\label{sec:interaction}

\begin{figure*}[t]
 \centering
 \includegraphics[width=\linewidth]{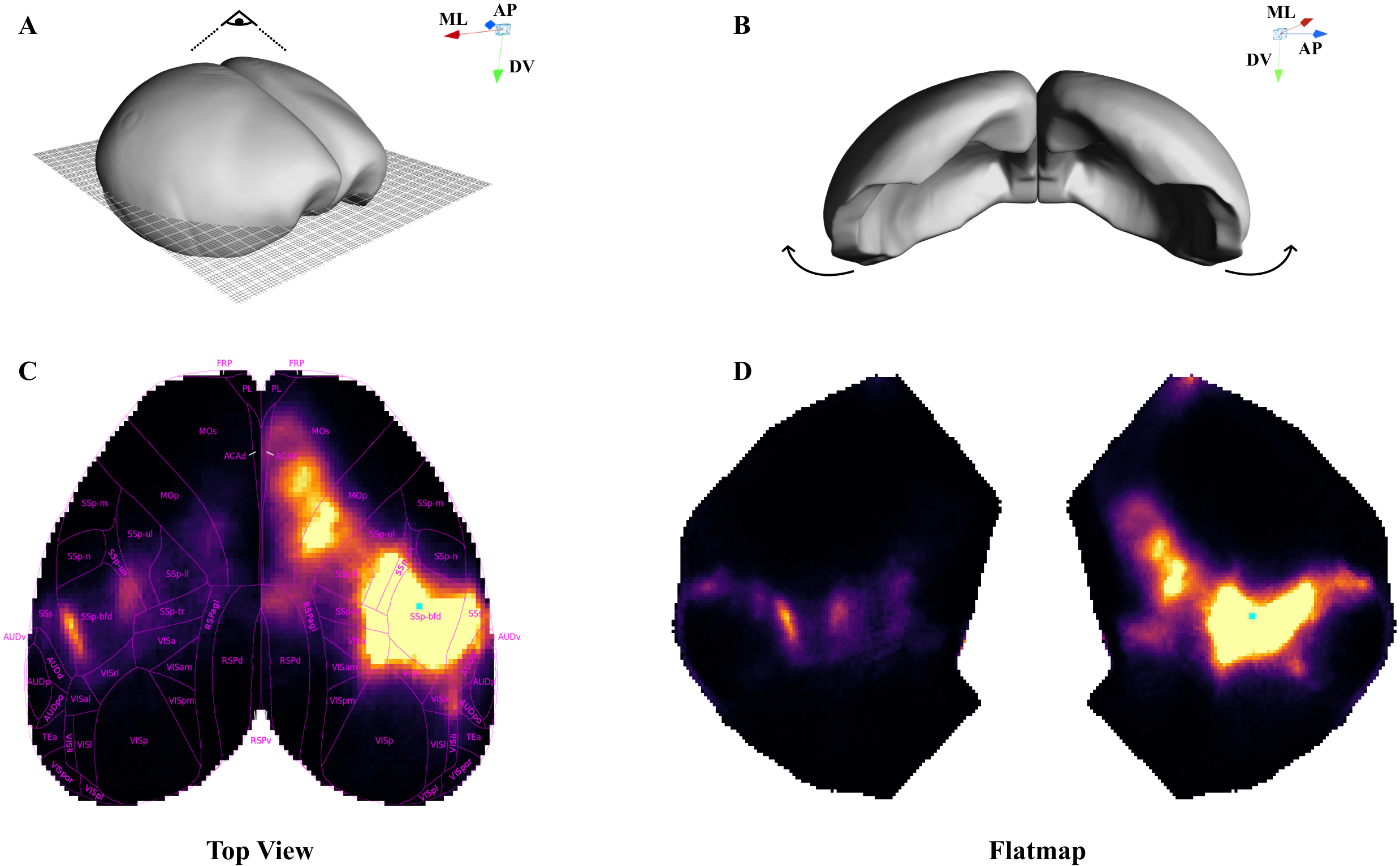}
  \caption{\textbf{A}
   Schematic of the top view perspective using the Allen BrainExplorer 3-D model of cortex \cite{lau_exploration_2008}.
   Axes are AP--anterior-posterior, ML--medial-lateral, and 
   DV--dorsal-ventral.
  \textbf{B} 
Schematic of the flatmap perspective which flattens the cortex, 
giving a better view of the regions curled beneath the cortex.
  \textbf{C}
  The top view projection with cortical regions overlaid in magenta. The source voxel (cyan) is seen in the SSp-bfd region. The projection strength is shown by the hot colors, where lighter shades reflect stronger connectivity and darker shades reflect weaker connectivity. 
  \textbf{D}
  The flatmap projection with the same source voxel as \figref{main}.
  Projections lateral to the injection site in somatosensory regions are more evident in this projection than the top view.
  }
  \label{fig:main}
\end{figure*}

BrainViewer can display estimated projections between a source and target voxels using either the top view or flatmap. 
The top view projection is shown in \figref{main}C including an overlay of the cortical regions. 
With this view the user sees projection strength between voxels in addition to context for each region.
For instance, when navigating the visual cortex, projections vary from source voxels within primary visual cortex (VISp), an area known to be important for retinotopy \cite{zhuang2017}.

The other projection BrainViewer uses is a cortical flatmap, shown in \figref{main}D. 
The flatmap is a truer representation of the curved surface of the cortex,
introducing less distortion than top view for the most medial areas \cite{vanessen2012a,vanessen2013,gamanut_mouse_2018}.
This provides information about voxels not directly seen from the top view projection as implied by \figref{main}B.

The user may click on any pixel within the right hemisphere of the cortex to select a source voxel for plotting \figref{gui}.
Once a source voxel is chosen, the arrow keys on the keyboard can precisely move about the brain, adjusting the source to an adjacent voxel. 
BrainViewer also has the ability to display images by following the cursor, instead of clicking a precise location.
However, this feature is limited by the speed at which plots can be generated. 
Moving the cursor quickly can result in lag, but the tool still performs well when moving slowly. 
The real-time plotting method preserves the source voxel when switching between flatmap and top view.
This allows users to see the differences in projections between the two views for a given injection.

\section{Methods}
\label{sec:methods}

\subsection{Data preparation and software}

BrainViewer is written in Python 3 \cite{harris2020array, 2020SciPy-NMeth, Hunter:2007} and uses the {\tt allensdk} package to access the data from the mouse brain connectivity atlas \cite{oh_mesoscale_2014}.
The voxel connectome package {\tt mcmodels} constructs the connectivity matrix  
\cite{knox_high-resolution_2018}.
The following equation describes the kernel smoothing regression model used to develop a low-rank connectivity weight matrix:
\begin{equation}
W_{ij} = 
\frac{\sum_{e} 
K(|| v_j - c_e||) Y_{ie}}
{\sum_{f} 
K(|| v_j - c_f||)}
\end{equation}
Here $W_{ij}$ is the projection strength from source voxel $j$
to target voxel $i$ expressed as a sum of the projection data $Y_{ie}$
of experiment $e$ weighted by a kernel function of its distance to the source;
see \cite{knox_high-resolution_2018} for more details.
The result is a large, dense but low-rank 
(rank of $W$ is bounded by the number of tracing experiments) 
matrix with $O(10^{11})$ entries containing connectivity strengths between 3-D voxels.
This matrix is developed from 428 experiments with 126 of these injections in the cortex.
The connection weight matrix underlying these data is large and cannot fit in memory. 
In Section~\ref{sec:optimizations} we describe some optimizations and 
precomputations that address this challenge.

\subsection{Brain projection views}
\label{sec:views}
Our tool allows users to switch between two views of the connectome, a task made possible by the ``mapper'' function from {\tt mcmodels}, 
which maps a 3-D volumetric image into a 2-D projection.
The mapper use a lookup table and path dictionary to define the geometry of the projections.
The lookup dictionary maps $(x_p,y_p)$ 
coordinates in the 2-D projection to a path through the 3-D volume. 
The index returned by the lookup allows our tool to reference positional data in the paths dictionary for the user-defined voxel coordinate and draw a new plot. Additionally, the lookup and paths dictionaries allow the BrainViewer tool to transition between top view and flatmap projections while maintaining the same source voxel. 
The paths dictionary returns an array of $(x, y, z)$ triples, representing a vectored path of voxels through the cortex, for the path ending at the corresponding location
in the lookup table.

For instance, if the user clicks on a voxel in VISa (\figref{gui}), 
we use their mouse coordinates to get $(x_p,y_p)$ in the projection,
then the lookup table tells us which voxels $\{(x_i, y_i, z_i)\}_i$ 
lie underneath. 
We plot the connectivity from the voxel in the middle of this path, corresponding to one of the middle cortical layers.
\figref{main}A shows how the top view is reflective of a top-down view of a three-dimensional model, giving a $114\times132$ bird’s eye view of the brain directly corresponding to 
medial-lateral (ML) $\times$ anterior-posterior (AP) axes.
An example of such data for a source voxel in SSp-bfd (primary somatosensory barrel field, the whisker sensory region) 
is shown in \figref{main}C.
The top view helps users contextualize the plotted connectivity for a given source voxel with a cortical region overlay.

Our tool also includes a flatmap of the connectome, seen in \figref{main}B. 
The flatmap is projection flattens the curved cortex by following paths which actually trace cortical depth, rather than just using the dorsal-ventral axis. 
The flatmap's $272\times136$ coordinates roughly correspond to a curved set
of ML-AP axes.
This reveals the complete cortical surface, including areas which are curled underneath and inaccessible in the top view. 
For the example shown in \figref{main}D, projections from the same voxel
in SSp-bfd that extend more deeply into adjacent lateral areas are more evident than in the top view (compare \figref{main}C).

Because all top view points are visible in the flatmap, there is no issue switching from top view to the flat map projection while preserving the source voxel. 
However, switching from flatmap to top view has the problem that some points shown in flatmap are not visible in the top view perspective.
In this case, our method computes the closest source voxel in the top view projection.

\subsection{Optimizations for speed and portability}
\label{sec:optimizations}
We have developed two versions of BrainViewer which function similarly but use different methods under-the-hood to generate the images.
The first method implements real-time plot generation from a $(x_p, y_p)$ coordinate pair after a mouse click event. 
Using these coordinates and the mapper, the corresponding column vector in the projection matrix $W$ is computed and plotted.
The entire low-rank decomposition of $W$ must be loaded at runtime along with the mapper utilities. 
These steps are not particularly efficient, leading to a nearly two minute preprocessing before any data can be displayed and a noticeable delay between image refreshes after the user clicks.

To speed up this process and also enable greater portability across datasets, 
we built a version of BrainViewer which can work directly with image stacks in the projected coordinate system.
We preprocess the projection plots as images that are named by the $(x_p, y_p)$ coordinates of the source voxel being displayed. 
Because we do not need to load the $W$ matrix or compute any 2-D projections, this version of BrainViewer starts quickly and has a much faster refresh latency.
There are only two requirements for formatting images from a new dataset: 
images are named based on the source voxel coordinates, and the coordinate space of the connectome must be easily mapped to the coordinate space of the image that the user is viewing and interacting with. 
For us, this is simply a scaling factor between the voxel coordinate space and the resolution of the saved images.

\section{Conclusions}
BrainViewer is a point-and-click software for viewing the projections from any voxel in a flattened view of the brain to any other voxel with high resolution and fast response.
The comprehensive spatial mouse cortical connectome is presented at the voxel level, which allows for more in depth analysis.
While we show results here for mouse, the framework is flexible and other datasets
can be incorporated easily.
Our code is freely available and documented online.

Using a GUI interface in Python allows us to overcome earlier limitations encountered by BrainModules \cite{Hira_Modules},
which displays connectivity in a small window, making it challenging to view specific sections of the brain. 
However, BrainModules does incorporate other datasets using a correlation-based method to estimate the connectome, which is different than the smoothing method of \cite{knox_high-resolution_2018}. 
BrainModules also provides multiple volumetric projections simultaneously.

In the future, we hope that BrainViewer will be used to visualize the connectomes of other species where tracing data are being collected such as marmoset, rat, etc. 
These data are not as easily accessible as the Allen Institute's mouse dataset, and voxel-based connectomes in those species have not been published.
We would like to incorporate multiple projections as well as overlays in the flatmap projection.
Other desirable features include display of the projection data in a native 3-D interface and a web-based version.
Visualization software like BrainViewer will continue to be an important way to understand connectomic data in the future.

\acknowledgments{
Thank you to Riichiro Hira and Lydia Ng for 
inspiration and encouragement.
This work was a senior capstone project for the first three authors' bachelor's degrees in Computer Science from Western Washington University.
}

\bibliographystyle{abbrv-doi}

\bibliography{src}

\end{document}